\PassOptionsToPackage{unicode}{hyperref}
\PassOptionsToPackage{hyphens}{url}
\documentclass[
]{article}
\usepackage{amsmath,amssymb}
\usepackage{lmodern}
\usepackage{ifxetex,ifluatex}
\ifnum 0\ifxetex 1\fi\ifluatex 1\fi=0 
  \usepackage[T1]{fontenc}
  \usepackage[utf8]{inputenc}
  \usepackage{textcomp} 
\else 
  \usepackage{unicode-math}
  \defaultfontfeatures{Scale=MatchLowercase}
  \defaultfontfeatures[\rmfamily]{Ligatures=TeX,Scale=1}
\fi
\IfFileExists{upquote.sty}{\usepackage{upquote}}{}
\IfFileExists{microtype.sty}{
  \usepackage[]{microtype}
  \UseMicrotypeSet[protrusion]{basicmath} 
}{}
\makeatletter
\@ifundefined{KOMAClassName}{
  \IfFileExists{parskip.sty}{%
    \usepackage{parskip}
  }{
    \setlength{\parindent}{0pt}
    \setlength{\parskip}{6pt plus 2pt minus 1pt}}
}{
  \KOMAoptions{parskip=half}}
\makeatother
\usepackage{xcolor}
\IfFileExists{xurl.sty}{\usepackage{xurl}}{} 
\IfFileExists{bookmark.sty}{\usepackage{bookmark}}{\usepackage{hyperref}}
\hypersetup{
  pdftitle={Works-magnet: Accelerating Metadata Curation for Open Science},
  pdfkeywords={open science, metadata curation, affiliation
matching, scholarly communication},
  hidelinks,
  pdfcreator={LaTeX via pandoc}}
\usepackage[left=3cm, right=3cm, top=3cm, bottom=3cm]{geometry}
\usepackage{longtable,booktabs,array}
\usepackage{calc} 
\usepackage{etoolbox}
\makeatletter
\patchcmd\longtable{\par}{\if@noskipsec\mbox{}\fi\par}{}{}
\makeatother
\IfFileExists{footnotehyper.sty}{\usepackage{footnotehyper}}{\usepackage{footnote}}
\makesavenoteenv{longtable}
\usepackage{graphicx}
\makeatletter
\def\maxwidth{\ifdim\Gin@nat@width>\linewidth\linewidth\else\Gin@nat@width\fi}
\def\maxheight{\ifdim\Gin@nat@height>\textheight\textheight\else\Gin@nat@height\fi}
\makeatother
\setkeys{Gin}{width=\maxwidth,height=\maxheight,keepaspectratio}
\makeatletter
\def\fps@figure{htbp}
\makeatother
\ifluatex
  \usepackage{selnolig}  
\fi
\newlength{\cslhangindent}
\newlength{\csllabelwidth}
 {
  \setlength{\parindent}{0pt}
  \ifodd #1 \everypar{\setlength{\hangindent}{\cslhangindent}}\ignorespaces\fi
  \ifnum #2 > 0
  \setlength{\parskip}{#2\baselineskip}
  \fi
 }%
 {}
\usepackage{calc}

\newenvironment{cslreferences}%
  {\setlength{\parindent}{0pt}%
  \everypar{\setlength{\hangindent}{\cslhangindent}}\ignorespaces}%
  {\par}

\title{Works-magnet: Accelerating Metadata Curation for Open Science}
\usepackage{authblk}
\author[%
  1%
  ]{%
  Eric Jeangirard%
}
\affil[1]{French Ministry of Higher Education and Research, Paris,
France}
\date{June 2025}

\makeatletter
\def\@maketitle{%
  \newpage \null \vskip 2em
  \begin {center}%
    \let \footnote \thanks
         {\LARGE \@title \par}%
         \vskip 1.5em%
                {\large \lineskip .5em%
                  \begin {tabular}[t]{c}%
                    \@author
                  \end {tabular}\par}%
                                                \vskip 1em{\large \@date}%
  \end {center}%
  \par
  \vskip 1.5em}
\makeatother

\begin{document}
\maketitle
\begin{abstract}
The transition to Open Science necessitates robust and reliable
metadata. While national initiatives, such as the French Open Science
Monitor, aim to track this evolution using open data, reliance on
proprietary databases persists in many places. Open platforms like
OpenAlex still require significant human intervention for data accuracy.
This paper introduces Works-magnet, a project by the French Ministry of
Higher Education and Research (MESR) Data Science \& Engineering Team.
Works-magnet is designed to accelerate the curation of bibliographic and
research data metadata, particularly affiliations, by making automated
AI calculations visible and correctable. It addresses challenges related
to metadata heterogeneity, complex processing chains, and the need for
human curation in a diverse research landscape. The paper details
Works-magnet's concepts, and the observed limitations, while outlining
future directions for enhancing open metadata quality and reusability.
The works-magnet app is open source on github
https://github.com/dataesr/works-magnet
\end{abstract}


\hypertarget{introduction}{%
\section{1. Introduction}\label{introduction}}

Open Science has gained considerable momentum globally, emphasizing the
need for transparent, accessible, and reusable research outputs and
data. In France, the Ministry of Higher Education and Research (MESR)
launched the National Plan for Open Science in 2018, establishing a
suite of quantitative monitoring tools. Since 2019, these tools have
also been made available to institutions for local monitoring. However,
a significant challenge remains: most institutions still depend on
proprietary databases for listing their publications and datasets. This
dependence hinders the comprehensive and open monitoring of scientific
activity.

Beyond merely tracking Open Science, there is a critical and growing
need for free, high-quality information on research. Major platforms,
such as Web of Science, which are widely used for generating metrics and
evaluating researchers, are proprietary. While open bibliographic
databases like OpenAlex offer an alternative, they are not immune to
inaccuracies. For instance, in the French context in a research entity
can up to five or more supervisors, a substantial number of articles are
incorrectly assigned to their institutions within OpenAlex. This
highlights that, just like proprietary data, open data sources
necessitate human curation by experts to improve metadata quality,
especially for affiliations.

In response to these challenges, the MESR Data Science \& Engineering
Team developed Works-magnet (works-magnet.esr.gouv.fr), launched in
2024. Works-magnet is a crucial initiative aimed at accelerating
metadata curation to measure the evolution of Open Science in France
using reliable, open, and controlled data. It serves as a tool to aid in
spotting, curating affiliations in OpenAlex. More broadly preaking, the
Works-magnet can help curating metadata produced by AI and machine
learning algorithms, giving a tool to put humans back in the loop.
Works-magnet can also be used to curate dataset and software mentions
for example, or even grants and financial support metadata.

\hypertarget{challenges-in-open-bibliographic-data-curation}{%
\section{2. Challenges in Open Bibliographic Data
Curation}\label{challenges-in-open-bibliographic-data-curation}}

The landscape of research metadata is vast and complex, encompassing
both ``classical'' metadata and emerging new types. Classical metadata
includes author names (with disambiguation challenges), affiliations,
titles, abstracts, publication types, thematic classifications,
reference lists, and funding information. Newer metadata types extend to
include information on Article Processing Charges (APCs), utilized
research infrastructure, financial support beyond funded projects, and
the reuse or sharing of datasets, software, and clinical trials.

The collection and processing of such diverse metadata are further
complicated by the variety of sources and techniques involved. Data can
originate from publishers, archives, authors, funders, and aggregators
such as scanR and OpenAlex. The publication processing chain itself is
intricate, involving multiple steps: harvesting from various sources,
deduplication, identifying authors and their affiliations, tracking
citations, and classifying scientific themes.

A significant hurdle lies in the accuracy of automatically generated
metadata, particularly affiliations. While various third-party tools,
many employing machine learning and artificial intelligence, exist to
match organization names to Research Organization Registry (ROR) IDs,
their accuracy rates, even ranging from 85\% to 95\% before human
intervention, are not perfect. These tools, including OpenAlex
Institution Parsing, S2AFF, RORRetriever, EMBL-EBI ROR Predictor,
dataESR affiliation matcher (L'Hôte and Jeangirard 2021), AffilGood
matcher (Duran-Silva et al. 2024), and various ROR Predictors developed
by Adam Buttrick. NonThis necessitates a ``human in the loop'' approach.

Despite technical advancements, human curation remains essential, even
for seemingly ``simple'' metadata like affiliations. This is
particularly true in France, where the administrative landscape with
many supervisors, university mergers, and new types of public
institutions make the affiliation landscape difficult to track. Beyond
simple alignment, other frictions arise, such as incomplete or erroneous
national or international reference systems, and the inconsistent
availability of raw metadata (signatures) from different sources like
publisher data, web scraping, PDF parsing (e.g., Grobid), and open
archive metadata. Furthermore, there are challenges in managing links
with software, including implicit mentions and aligning software with
SWHIDs via URLs , and difficulties with research datasets, characterized
by Datacite indexing issues (where one DOI does not necessarily equal
one dataset) and high heterogeneity

\hypertarget{works-magnet-a-solution-for-open-metadata-curation}{%
\section{3. Works-magnet: A Solution for Open Metadata
Curation}\label{works-magnet-a-solution-for-open-metadata-curation}}

Works-magnet represents a strategic shift from proprietary to open
environments for research data curation. In a proprietary setting,
access is restricted to customers, discussions are private, and
corrected data often becomes proprietary, reinforcing dependence on the
original tools. In contrast, Works-magnet operates within an open
environment, accessible to any user, with transparent processes for
requesting corrections. The corrected data within this open framework
becomes open and reusable by anyone, leveraging the workforce of public
employees to enhance the quality of open data. This paradigm promotes an
open productivity tool to facilitate curation.

One of the fundamental shifts brought about by Works-magnet lies in its
paradigm change, moving from a proprietary data curation environment to
an open model. In a proprietary setting, access to tools and data is
typically limited to paying customers, discussions are private, and
metadata corrections, while improving quality, remain the property of
the platform, thus reinforcing user dependence. In contrast,
Works-magnet embodies a radically open approach. The system is
accessible to everyone, correction requests are managed transparently
via public platforms like GitHub, and crucially, the corrected data
becomes open and reusable resources for anyone. This model leverages the
contribution of the public sector workforce to enrich the quality of
open data, transforming curation into a collaborative and transparent
process that serves the entire scientific community and beyond.

\begin{longtable}[]{@{}ll@{}}
\toprule
\begin{minipage}[b]{0.47\columnwidth}\raggedright
\textbf{Proprietary environment}\strut
\end{minipage} & \begin{minipage}[b]{0.47\columnwidth}\raggedright
\textbf{Open environment}\strut
\end{minipage}\tabularnewline
\midrule
\endhead
\begin{minipage}[t]{0.47\columnwidth}\raggedright
reserved for customers\strut
\end{minipage} & \begin{minipage}[t]{0.47\columnwidth}\raggedright
open to any user \strut
\end{minipage}\tabularnewline
\begin{minipage}[t]{0.47\columnwidth}\raggedright
private discussions and data exchange\strut
\end{minipage} & \begin{minipage}[t]{0.47\columnwidth}\raggedright
transparency in requesting corrections\strut
\end{minipage}\tabularnewline
\begin{minipage}[t]{0.47\columnwidth}\raggedright
Excel? Database extracts?\strut
\end{minipage} & \begin{minipage}[t]{0.47\columnwidth}\raggedright
(open) productivity tool to facilitate curation\strut
\end{minipage}\tabularnewline
\begin{minipage}[t]{0.47\columnwidth}\raggedright
Corrected data becomes proprietary\strut
\end{minipage} & \begin{minipage}[t]{0.47\columnwidth}\raggedright
Corrected data is open and can be reused by anyone\strut
\end{minipage}\tabularnewline
\begin{minipage}[t]{0.47\columnwidth}\raggedright
reinforces dependence on proprietary tools\strut
\end{minipage} & \begin{minipage}[t]{0.47\columnwidth}\raggedright
workforce of public employees is used to improve open data quality\strut
\end{minipage}\tabularnewline
\bottomrule
\end{longtable}

\begin{figure}
\centering
\includegraphics[width=6.25in,height=\textheight]{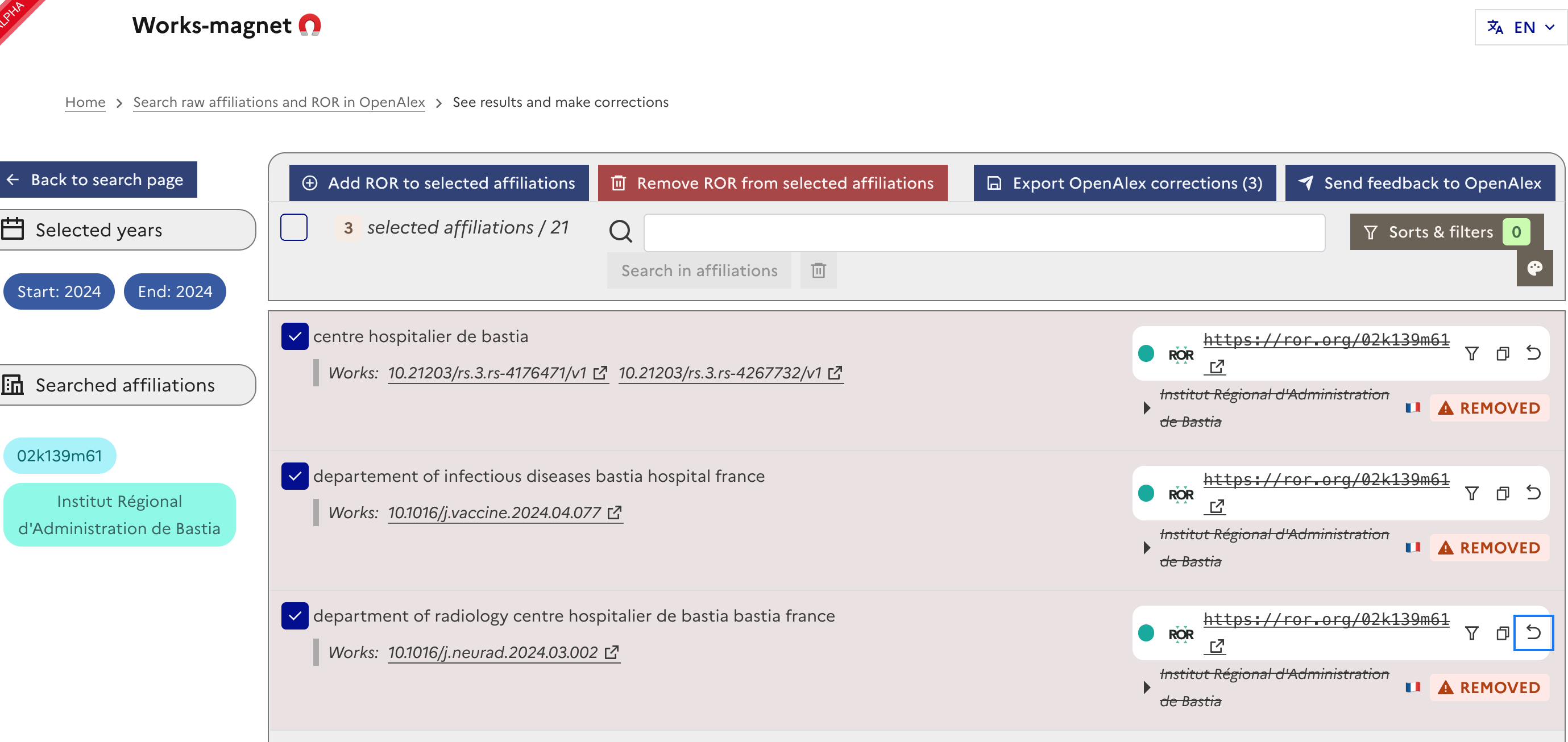}
\caption{Works-magnet screenshot}
\end{figure}

\hypertarget{code-and-data-availibility}{%
\section{4. Code and data
availibility}\label{code-and-data-availibility}}

The works-magnet app itself is open source on GitHub
https://github.com/dataesr/works-magnet

The curated data produced is available both on GitHub via issues
https://github.com/dataesr/openalex-affiliations/issues and with open
dataset
https://data.enseignementsup-recherche.gouv.fr/explore/dataset/openalex-affiliations-corrections/table/

\hypertarget{monitoring-and-limitations}{%
\section{5. Monitoring and
Limitations}\label{monitoring-and-limitations}}

The efficacy and ongoing development of Works-magnet are supported by
robust monitoring of correction requests and transparent reporting.
Correction requests are systematically tracked, primarily through
platforms like GitHub issues. As of recently, 71,283 corrections had
been requested, with a significant proportion already closed. Data on
these corrections, including their status (open or closed) and the
domain of the top contributors, is publicly available. This transparency
allows for a clear overview of the ongoing curation efforts and helps
identify areas requiring more attention. Specific dashboards are also
available to track correction requests for individual establishments.

Despite its innovative approach, Works-magnet faces several limitations.
Technical challenges include reliance on the GitHub API, and delays in
verification by OpenAlex, which can accumulate a backlog of corrections.
The completeness and maintenance of the ROR registry itself also pose a
challenge, although many national projects and their synchronization
with ROR are attempting to address this. A persistent issue is the
inconsistent availability of raw signatures (affiliation strings as they
appear in publications), necessitating diverse strategies like web
scraping, PDF parsing (e.g., Grobid), and extracting metadata from open
archives.

Resource constraints are a significant hurdle for Works-magnet. The
project operates with virtually no financial sponsorship and minimal
human resources, with less than 0.25 Full-Time Equivalent (FTE)
allocated per year. This limited capacity affects the pace of
development and the ability to address all identified issues promptly.

Furthermore, specific difficulties are encountered when linking research
outputs with software and datasets. For software, challenges include the
presence of implicit mentions within texts and the complexities of
aligning software with Software Heritage Identifiers (SWHIDs) via URLs.
For research datasets, the primary difficulties are related to Datacite
indexing, where one DOI does not always correspond to a single dataset,
and the inherent high heterogeneity of research datasets themselves.

\hypertarget{future-directions}{%
\section{6. Future Directions}\label{future-directions}}

The future development of Works-magnet and the broader open metadata
ecosystem holds several promising directions. A key objective is to
ensure that the results of metadata curation are not only open but also
interoperable, facilitating their reuse in diverse contexts beyond the
initial scope of Works-magnet. This would maximize the value of the
curated data for the entire research community.

Furthermore, the continuously growing and refined dataset resulting from
Works-magnet's curation efforts has the potential to serve as a valuable
training base for new artificial intelligence models. This could lead to
the development of more accurate and efficient automated curation tools
in the future, potentially reducing the reliance on extensive human
intervention for certain tasks (Jeangirard 2022).

Finally, there is an ongoing discussion about centralizing the results
of various curation initiatives to simplify their dissemination. Such a
centralized approach could create a single, authoritative source for
high-quality, openly curated research metadata, further advancing the
goals of Open Science by making reliable data readily available to all
stakeholders. These future directions underscore the commitment to
building a sustainable and comprehensive open metadata infrastructure.

\hypertarget{references}{%
\section*{References}\label{references}}
\addcontentsline{toc}{section}{References}

\hypertarget{refs}{}
\begin{cslreferences}
\leavevmode\hypertarget{ref-duran-silva-etal-2024-affilgood}{}%
Duran-Silva, Nicolau, Pablo Accuosto, Piotr Przybyła, and Horacio
Saggion. 2024. ``AffilGood: Building Reliable Institution Name
Disambiguation Tools to Improve Scientific Literature Analysis.'' In
\emph{Proceedings of the Fourth Workshop on Scholarly Document
Processing (Sdp 2024)}, edited by Tirthankar Ghosal, Amanpreet Singh,
Anita Waard, Philipp Mayr, Aakanksha Naik, Orion Weller, Yoonjoo Lee,
Shannon Shen, and Yanxia Qin, 135--44. Bangkok, Thailand: Association
for Computational Linguistics.
\url{https://aclanthology.org/2024.sdp-1.13/}.

\leavevmode\hypertarget{ref-jeangirard:hal-03819060}{}%
Jeangirard, Éric. 2022. ``L'utilisation de l'apprentissage automatique
dans le Baromètre de la science ouverte~: une façon de réconcilier
bibliométrie et science ouverte~?'' \emph{Arabesques}, no. 107
(September): 10--11. \url{https://doi.org/10.35562/arabesques.3084}.

\leavevmode\hypertarget{ref-lhote_using_2021}{}%
L'Hôte, Anne, and Eric Jeangirard. 2021. ``Using Elasticsearch for
Entity Recognition in Affiliation Disambiguation.''
\emph{arXiv:2110.01958 {[}Cs{]}}, October.
\url{http://arxiv.org/abs/2110.01958}.
\end{cslreferences}

\end{document}